# MINIATURISED SPOTTER-COMPATIBLE MULTICAPILLARY STAMPING TOOL FOR MICROARRAY PRINTING

Alexei L Drobyshev[1], Nikolai N Verkhodanov[1], Alexander S Zasedatelev[1,2]

[1]Engelhardt Institute of Molecular Biology, Moscow, and [2]Moscow Institute of Physics and Technology, Dolgoprundy, Russian Federation

Address correspondence to:
Alexei L Drobyshev
Engelhardt Institute of Molecular Biology
Vavilov str. 32, Moscow, 119991
Russian Federation
dro@newmail.ru

ABSTRACT
Novel microstamping tool for microarray printing is proposed. The tool is capable to spot up to 127 droplets of different solutions in single touch. It is easily compatible with commercially available microarray spotters. The tool is based on multichannel funnel with polypropylene capillaries inserted into its channels. Superior flexibility is achieved by ability to replace any printing capillary of the tool. As a practical implementation, hydrogel-based microarrays were stamped and successfully applied to identify the *Mycobacterium tuberculosis* drug resistance.

INTRODUCTION
Microarray technology is the most successful example of miniaturisation in modern life science. Taking advantage of high parallelism resulting form miniaturisation microarrays has become a tool of choice in numerous applications (1). However in contrast to microarrays themselves, the methods of their manufacturing are generally completely macroscopic.
Currently the most of microarrays are produced by pin or piezo-jet technology. Both use 384 (seldom 96) well micrtotiter plates filled with the solutions to be spotted. Pins or piezo dispensers are placed in the holder so that they match the pattern of plate, i.e. in case of 384 well plate they have 4.5 mm pitch. This imposes some restrictions on the shape, size and density of arrays produced in multipin mode. These arrays consist of clusters of spots matching the pattern of the microtiter plate they were printed from. For example, no microarray smaller than 4.5 by 4.5 mm can be printed in multipin mode. If a high density microarray of hundred spots is desired it can be produced in single pin mode only. On the other hand in highly parallel mode with 48 (typically 4 by 12) pins, the resultant array will cover entire 25×75 mm microscopic slide, i.e. high density arrays of small or middle size cannot be spotted in highly parallel mode.
The best known exception from this general trend is the method of *in situ* synthesis of microarray features like light-directed oligonucleotide synthesis (2) of Affymetrix platform. Regarded by many as the most sophisticated platform, it requires industrial environment for microarray manufacturing; it has strict limitation on the length of synthesised products (generally no more than 25 nucleotides) and cannot be used for arrays of antibodies, living cells, etc.
The less known examples are based on parallel depositing of pre-synthesised species by contact (3-6) or non-contact (7-10) method. Contact system exploits a microfluidic chip with array of 144 nozzles connected to 8 reservoirs (18 nozzles per one reservoir) at the opposite side of the chip. Reservoirs and corresponding nozzles are filled with spotting solutions. Then the chip is pressed to the printing surface "for less than a few minutes" (5) leaving spots on it. In non-contact system the array contains 24 or 96 nozzles. Every nozzle is connected through a microchannel with its reservoir (one nozzle per reservoir). The pressure pulse is applied simultaneously to all the reservoirs causing every nozzle to shoot the droplet. These methods, however, require special machinery and are not easily compatible with existing microarray spotters available in many biological laboratories.
Here we propose a spotter-compatible microstamping tool capable to print up to 127 spots within 3×3 mm area in single touch. The tool is easy to assemble in general biological laboratory with the aid of institute workshop. With this tool we stamp a 70-element high density hydrogel-based microarray and use it for identification of drug-resistant strains of *Mycobacterium tuberculosis* (11).

MATERIALS AND METHODS
Multichannel funnels were purchased form TEGS (Saratov, Russian Federation), as multichannel holders of optical fibres. Polypropylene capillaries were drawn manually from BD Micro-Fine™ 1 ml syringes (Becton Dickinson, USA). Every capillary was inserted into the funnel and adjusted so that its outer diameter at the narrow end of funnel (monitored by means of optical microscope) was from 0.15 to 0.18 mm and then trimmed at this position by a sharp blade (Stanley Works, England). The blade sharpness is essential as it is a major factor affecting the quality of printing end of capillary, so the blades should be replaced as soon as necessary. At the opposite end every capillary was trimmed to achieve the necessary length (from 65 to 66 mm, depending on its position in the funnel).

The holder for multicapillary funnel was manufactured at the Institute workshop according to the Supplementary Figure S1.

*Mycobacterium tuberculosis* DNA samples were obtained from Moscow Antituberculosis Centre; their sequences were analysed by microarray hybridisation and direct sequencing earlier (11). The oligonucleotide-containing spotting solutions for hydrogel-based Multi-Drug Resistance TB-Biochip (11) were purchased from Biochip-IMB Ltd (Moscow, RF). The composition of solutions was described earlier (12). Additionally every solution contained 10μM of Texas Red™ dye (Molecular Probes) for evaluation of gel pad size upon the polymerisation. Slides treatment before and after spotting was performed as described earlier (12). DNA target preparation, hybridisation and subsequent washing of microarrays were done according to TB-Biochip-MDR protocol (11). Fluorescent images were acquired by Chip-Detector-01 and quantified by ImaGeWare version 2.38 (both Biochip-IMB, RF). For correlation plots mean fluorescent intensities were calculated for circle areas of fixed diameter snapped to the microarray features and the local fluorescent background was subtracted. Statistical analysis of the data was performed with Microsoft Excel and Origin 6.1 (Origin Lab, USA)

RESULTS AND DISCUSSION

The key element of our printing tool is a glass multicapillary funnel with 127 microchannels forming hexagonal array, Figure 1 (A). The funnel was manufactured at industrial environment by forming hexagonal array of glass tubes and baking them together. Then this multichannel structure was drawn in a special oven to obtain the funnel of desired shape. This technology was used earlier for e.g. manufacturing of polycapillary lenses (13).

At the broad end of funnel the diameter of microchannels is 1 mm with 1.3 mm pitch, at the narrow end the diameter is 0.21 mm with 0.27 mm pitch, and the funnel's length is 64.5 mm. These microchannels serve as guides for polypropylene capillaries used for printing, Figure 1 (B) and (C). The capillaries are filled with spotting solutions and inserted into microchannels so that they protrude from the narrow end of funnel by approximately 0.5 mm. At the broad end of funnel the capillaries are blocked from moving out by upper clamp (Figure 1 (A)) which is normally tightly attached to the funnel and removed only to access the capillaries. This makes the capillary working like a spring: when it touches the surface of substrate with its protruding end it shrinks exerting some pressure and forming tight contact necessary to transfer the liquid to the surface.

The entire assembly was placed into Genetix QArray spotter (by means of specially designed holder, Supplementary Figure S1) instead of standard pin tool. In the course of printing this stamp was brought in contact with slides so that it stamped the pattern formed by array of microcapillaries (Figure 1 (D)).

In order to test the efficiency of stamped arrays in mutation screening by hybridisation analysis the funnel was loaded with 70 capillaries filled with hydrogel-based spotting solution with oligonucleotide probes for *Mycobacterium tuberculosis* multi-drug resistance set (11) and 10 μM Texas Red™ fluorescent dye. Then the funnel was loaded into QArray spotter and a 70-element array was stamped to 10 slides in triplicate with 900 ms contact time. After spotting the tool was stored for 10 days at +4°C and successfully used for another spotting again (as demonstrated below).

Spotted slides were polymerised (12) and Texas Red™ fluorescence images were acquired and quantified. Relative standard deviation of integral signals from 70 elements was from 21 to 23% for different slides.

Then the microarrays were hybridised with wild type (H37Rv strain) and mutant DNA target (Figure 2A and B, respectively). Every microarray contained two replicates hybridised in the same volume (only one replicate shown at Figure 2). Clear difference between wild type and mutant hybridisation patterns was observed for microarray features corresponding to 531$^{st}$ codon of *rpoB* and 315$^{th}$ codon of *katG* genes in both replicates (selected area at Figure 2). Each group contains features with oligonucleotides matching the same position of target DNA but complementary to wild type and all known mutant sequences of particular codon (4 and 7 different mutations for 531$^{st}$ codon of *rpoB* and 315$^{th}$ codon of *katG*, respectively). For wild type DNA the strongest signals in both groups are produced by wild type oligonucleotides (first from the left in every group). For mutant DNA the strongest signal in *rpoB* 531 group corresponds to substitution of Ser to Leu, for *katG* 315 group it corresponds to Ser to Thr substitution, in agreement with earlier direct sequencing and hybridisation analysis (11). Separately the same pair of hybridisations was performed with the microarrays stamped after 10 days storage of the tool at +4°C. Both wild type and mutant DNA targets were successfully identified (not shown). This proves that our stamping tool provides sufficient printing accuracy for correct detection of single nucleotide substitutions in *Mycobacterium tuberculosis* DNA even without internal control like dual colour hybridisation.

To assess the effect of 10 days storage of the tool quantitatively two microarrays were hybridised against wild type DNA target and the hybridisation signals were plotted against that of array from Figure 2 (A): the first microarray was stamped in the same batch with the ones from Figure 2 and second microarray was stamped after the storage (Figure 3 (A), solid triangles and open squares, respectively). The correlation coefficients for the data before and after storage were 0.94 and 0.95, respectively. To assess the source of this variability, a correlation plot of two replicates hybridised in the same volume (the array of Figure 2 (A)) was made (Figure 3 (B)). The correlation coefficient was 0.993 indicating that the error coming from stamping is much smaller than the error from individual hybridisations or 10 days storage of loaded tool at +4°C.

As the data from Figure 3 (B) demonstrate a good reproducibility of droplets spotted by the same capillary, this reproducibility was assessed in a special experiment. The funnel was loaded with only one capillary containing hydrogel-based spotting solution with Cy5-labelled oligonucleotide marker from TB-Biochip-MDR set and placed into QArray spotter. An array of 1600 droplets was spotted with this capillary and polymerised; then Cy5 fluorescence image was acquired and quantified. Relative standard deviation of integral signals from randomly selected 7 by 10 massive of 70 spots was 6%.

Another important question of tool validation is the maximal number of replicates that could be stamped without refilling the capillaries. This number obviously depends on the initial loading of capillaries and the average volume of spots. The latter depends on the capillary diameter and the wetting of substrate by spotting solution. To study this second factor only one capillary was left in the funnel and 3 different solutions were used for spotting: the solution for hydrogel-based microarrays (12) (50% glycerol, 4.75% methacrylamide, 0.25% methylenebisacrylamide and Cy5-labelled marker oligonucleotide with amino group from TB-Biochip-MDR set in 0.01M Na–borate buffer, pH 10.5), 3M betaine with 25% glycerol and 4M betaine. Initial loading of capillary was adjusted to be maximal with the resulting spots diameter smaller than 0.27 mm (the pitch of printing end of funnel). Then the funnel was placed into QArray spotter and the routine for printing of 5000-element (50×100) microarray with 0.27 mm pitch and 900 ms contact time was started in single capillary mode on different slides from different batches. For hydrogel-based spotting solution four maximal achieved numbers of spots were 4150, 4050, 2850 and 2650 (average is 3425); for 25% glycerol and 3M betaine these numbers were 2500, 2400, 1650 and 1250 (average is 1950); for 4M betaine these were 3400, 1850, 1200 and 1100 (average is 1900, rounded to the nearest multiple of 50). The average maximal number of spots for hydrogel-based solution is significantly higher than this number for 3M betaine with 25% glycerol and 4M betaine (p-values are 0.025 and 0.059, respectively), in agreement with the suggestion that the transferred volume decreases with the decrease of wetting of substrate with spotting solution. Nevertheless, several hundreds of replicates are confidently achievable for any of tested solutions (e.g. for N=500 p-values are 0.005, 0.017 and 0.079 respectively).

The proposed method of microarray printing combines simplicity and flexibility because of ability to add remove or replace capillaries in the funnel if necessary. This ability could be used to achieve desired feature size distribution across the array. If desired, better spot size uniformity could be also achieved by imposing more stringent restrictions on the diameters of capillaries and adjusting proper speed of tool lifting after the stamping (which is impossible in current QArray interface). The stamping array could be easily expanded beyond current 127 elements. Since the funnel is 3-dimentional structure this expansion will not require any change of interconnection layout necessary for planar chip-based microstamping tools.

Our tool was found to be suitable for stamping of different solutions but it could stamp more replicates if the wetting of substrate is low, e.g. due to the presence of glycerol. After spotting the tool could be stored and used again without recharging the capillaries. In this case it is important that the solution do not dry or crystallise e.g. because of presence of betaine (14) or glycerol (12). With QArray spotter it takes no more than 3 seconds per stamping of up to 127 features, no washing, drying or refilling are required making our multicapillary funnel stamp a powerful tool for manufacturing of high density microarrays of small or middle size.

ACKNOLEDGEMENTS
We are grateful to N. Blow, V.M. Mikhailovich, N.V. Zakharova, O.A. Zasedateleva, E.N. Timofeyev, A.V. Chudinov, S.A. Pan'kov, for helpful discussion, to O.G. Somova and O.V. Moiseyeva for post-spotting microarray processing, to O.V. Markova, A.Y. Kozlova, E.E. Fesenko and O.V. Antonova for the microarray hybridisations.

COMPETING INTERESTS STATEMENT
Authors declare no competing interests

FIGURES

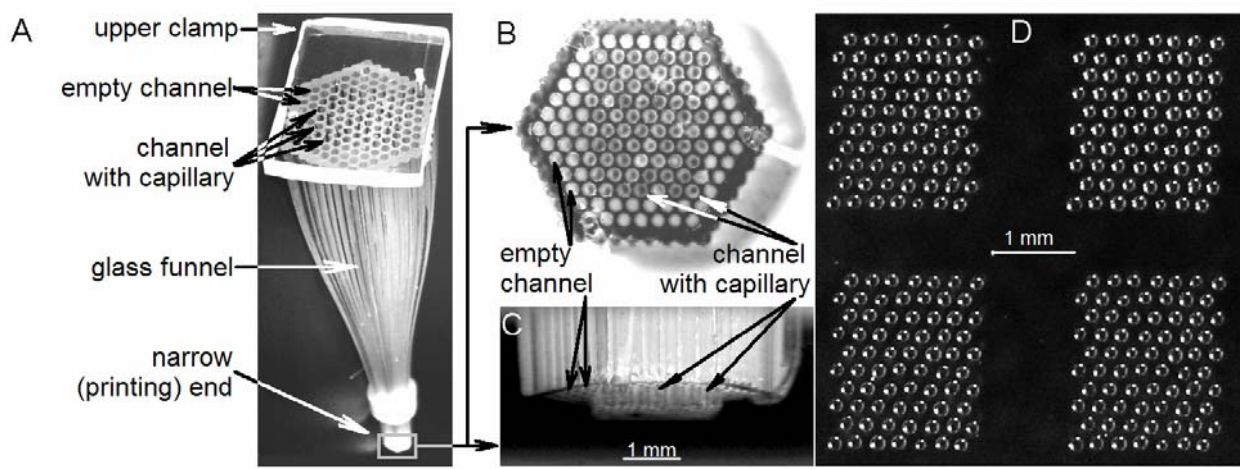

**Figure 1.** Multicapillary printing tool and spotted arrays. (A) General view of assembly: glass funnel with 127 microchannels, 70 of them have inserted polypropylene capillaries filled with solution to be spotted, the remaining 57 channels are not used for spotting. The upper clamp is normally tightly fixed at the broad end of funnel blocking the capillaries from moving out in upward direction. (B) Front view of narrow (printing) end of funnel with 70 channels having inserted capillaries and empty channels. (C) Side view of narrow end of funnel with 70 polypropylene capillaries protruding out of cannels. At figures (B) and (C) the arrows point to the same channels. (D) Microphotograph of 70-element microarrays of 4M betaine droplets printed with 70-capillary printing tool in single touch in quadruplicate.

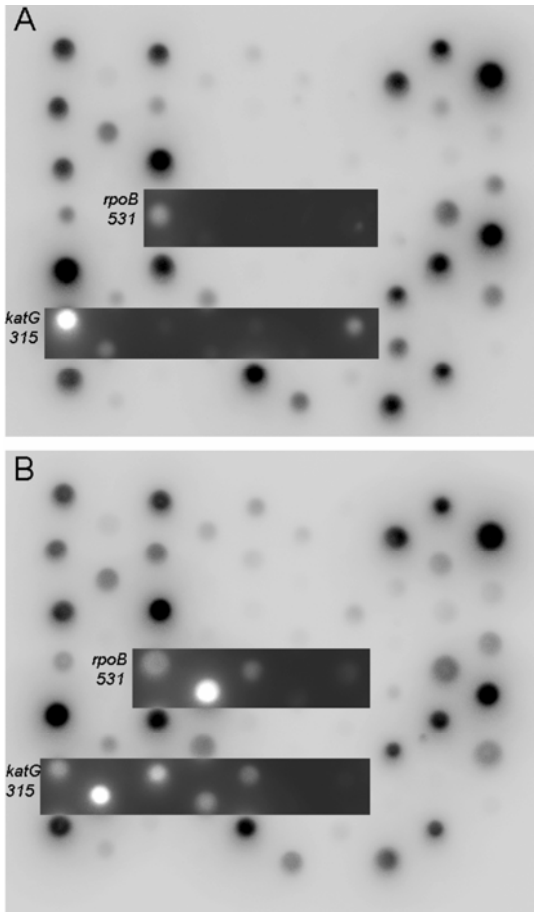

**Figure 2.** Analysis of *Micobacteriun tuberculosis* drug suspectability by hybridisation on stamped microarrays. (A) Wild type DNA sample. (B) Double mutant DNA sample: Ser531 > Leu (TCG > TTG) in the *rpoB* gene, Ser315 > Thr (AGC > ACC) in the *katG* gene.

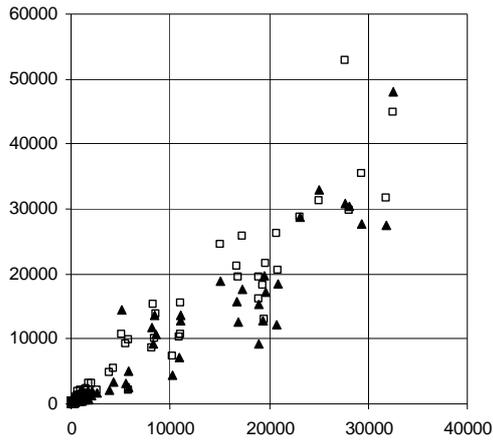

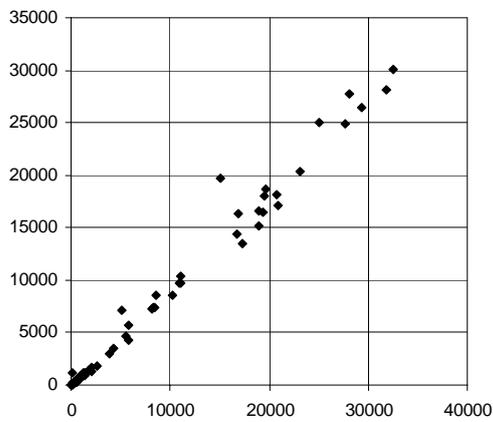

**Figure 3.** Correlations of hybridisation signals of wild type DNA target in different experiments. (A) Correlation of hybridisation signals for microarrays hybridised separately. Solid triangles: microarrays from the same batch. Open squares: microarray from the first batch versus microarray from the batch stamped after 10 days storage of the tool at +4°C. (B) Correlation of hybridisation signals of two microarray replicates stamped on the same slide and hybridised in the same volume. For both plots fluorescent background was subtracted from mean fluorescent intensities of microarray features.

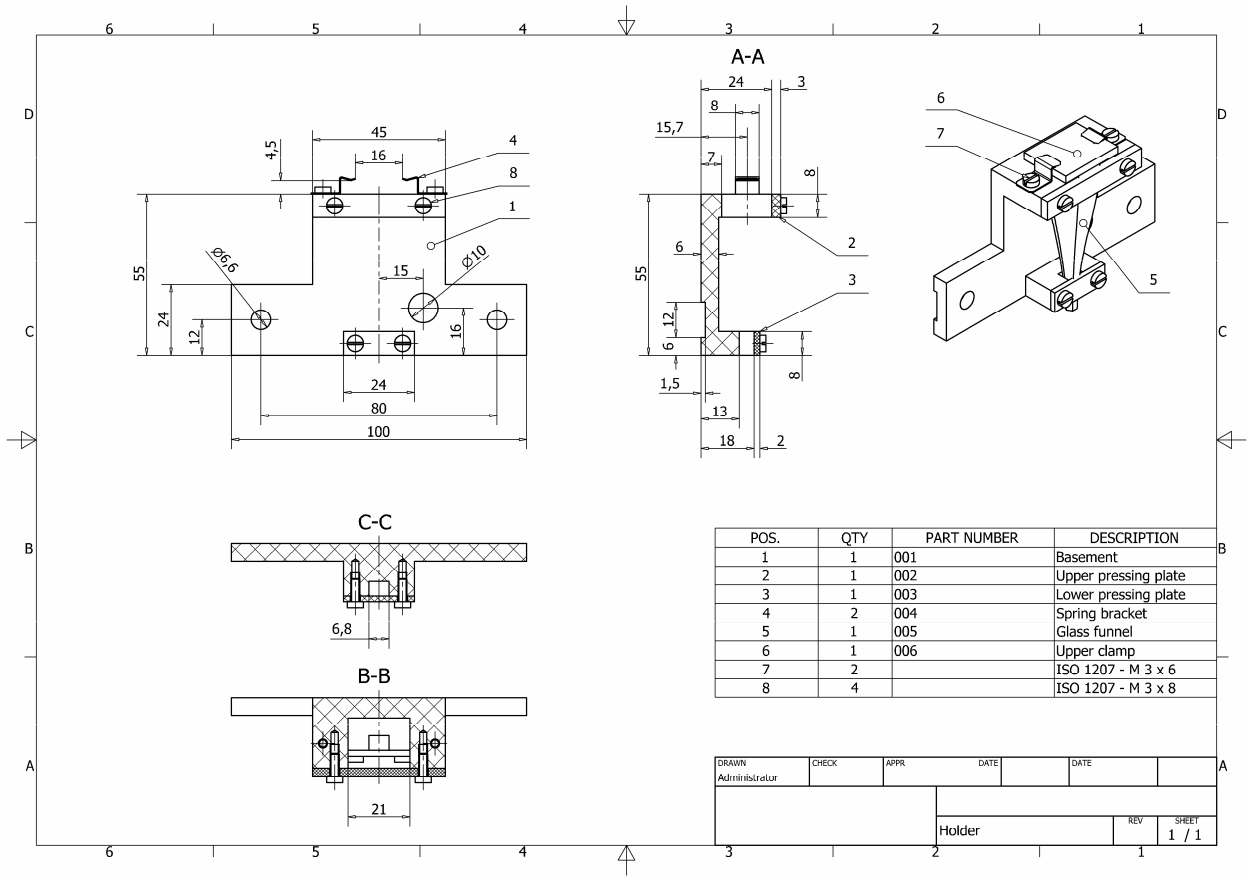

**Figure S1**